\documentclass[12pt,preprint]{emulateapj}

\begin{document}

\title{Molecular Hydrogen Formation on Amorphous Silicates \\
Under Interstellar Conditions }

\author{H.B. Perets\altaffilmark{1}, 
A. Lederhendler\altaffilmark{2},
O. Biham\altaffilmark{2}, 
G. Vidali\altaffilmark{3},
L. Li\altaffilmark{3}, 
S. Swords\altaffilmark{3}, 
E. Congiu\altaffilmark{3,4,5},
J. Roser\altaffilmark{6}, 
G. Manic\'o\altaffilmark{5}, 
J.R. Brucato\altaffilmark{7} 
and 
V. Pirronello\altaffilmark{5} 
}

\altaffiltext{1}{Faculty of Physics, Weizmann Institue of Science, Rehovot 76100, Israel} 
\altaffiltext{2}{Racah Institute of Physics, The Hebrew University, Jerusalem 91904, Israel} 
\altaffiltext{3}{Physics Department, Syracuse University, Syracuse, NY 13244, USA} 
\altaffiltext{4}{Universit\'a di Cagliari, Dipartimento di Fisica, Cagliari Italy}
\altaffiltext{5}{Universit\'a di Catania, DMFCI, 95125 Catania, Sicily, Italy} 
\altaffiltext{6}{NASA Ames, Mail Stop 245-6, Moffett Field, CA, 94035, USA} 
\altaffiltext{7}{INAF-Osservatorio Astronomico di Capodimonte, Napoli, Italy}

\begin{abstract}

Experimental results on the formation of molecular hydrogen on amorphous 
silicate surfaces are 
presented for the first time and
analyzed using a rate equation model. 
The energy barriers for the relevant diffusion and
desorption processes are obtained. 
They turn out to be significantly higher than those obtained 
earlier for polycrystalline silicates,
demonstrating the importance of grain morphology. 
Using these barriers we evaluate the 
efficiency of molecular hydrogen formation on amorphous silicate
grains under interstellar conditions. 
It is found that unlike polycrystalline silicates, amorphous silicate
grains are efficient catalysts of H$_{2}$ formation within a temperature
range which is relevant to diffuse interstellar clouds.
The results also indicate that the hydrogen molecules 
are thermalized with the surface
and desorb with low kinetic energy.
Thus, they are unlikely to occupy highly excited states. 

\end{abstract}


\section{Introduction}
\label{intro.}

H$_{2}$ is the most abundant molecule in the interstellar medium (ISM). 
It plays a crucial role in the initial cooling of clouds during 
gravitational collapse and is involved in most reaction schemes
that produce other molecules 
\citep{Duley1984}. 
It is widely accepted that 
H$_{2}$ formation in the ISM takes place on dust grains 
\citep{Gould1963}.
In this process, 
H atoms that collide with a grain and adsorb on its surface
quickly equilibrate and diffuse
either by thermal activation or tunneling.
They may encounter each other and form
H$_{2}$ molecules
\citep{Williams1968,Hollenbach1970}, 
or desorb thermally
in atomic form.

In recent years, we have conducted a series of 
experiments on the formation of molecular hydrogen on
dust grain analogues 
such as polycrystalline silicates 
\citep{Pirronello1997a,Pirronello1997b}, 
amorphous carbon
\citep{Pirronello1999}
and amorphous water ice
\citep{Manico2001,Roser2002,Roser2003,Perets2005},
under astrophysically relevant conditions.
In these experiments, the surface was irradiated by beams of H and D atoms.
The production of HD molecules 
was measured during the irradiation and during
a subsequent temperature programmed desorption (TPD) experiment. 
To disentangle the process of diffusion from the one of desorption,
separate experiments were carried out in which molecular species were
irradiated on the sample and were later induced to desorb. 
Related studies were done on 
amorphous ice surfaces 
\citep{Hornekaer2003,Hornekaer2005,Dulieu2005,Perets2005,Amiaud2006,Williams2007}. 

The results were analyzed using rate equation models 
and the energy barriers for the relevant diffusion and desorption
processes were obtained
\citep{Katz1999,Cazaux2004,Perets2005}.
Using these parameters, 
the conditions for efficient H$_{2}$ 
formation on different astrophysically relevant surfaces
were found. 
In particular, the
formation of H$_{2}$ on polycrystalline silicates 
was found to be efficient only in 
a narrow temperature window below 10K.
Since the typical dust grain temperature in diffuse interstellar
clouds is higher than 10K, 
these results
indicated that polycrystalline silicate grains cannot be efficient 
catalysts for H$_{2}$ formation in most diffuse clouds. 

In this \emph{Letter} we present, 
for the first time, 
experiments on molecular hydrogen formation 
on \emph{amorphous} silicates 
and analyze the results using a suitable rate equation model
\cite{Perets2005}. 
Using the parameters that best fit the experimental
results, the efficiency of hydrogen recombination on grains
is obtained for 
a range of conditions pertinent to diffuse interstellar clouds. 
It is found that
unlike the polycrystalline silicate grains,
amorphous silicate grains, which are the main silicate
component in interstellar clouds
\cite{Tielens2005},
are efficient catalysts for H$_{2}$ formation within a broad temperature
window that extends at least up to about 14K.

\section{Experimental Methods and Results}

\label{sec:Experimental} 

The apparatus consists of an ultra-high vacuum 
chamber housing the sample holder and a detector. 
The sample can be cooled by liquid helium 
to $\sim$5K, as measured by a calibrated
silicon diode and thermocouple placed 
in the back of the sample. 
A heater in the back of the sample is used to maintain 
a set temperature between 5 and 30K 
during the irradiation phase of the experiment. 
The sample and detector can rotate around the vertical axis. 
Prior to a series of measurements, the sample is heated to 380-400K. 
During a series of measurements, the sample is taken periodically to 
200-250K to desorb condensables. 
Hydrogen and deuterium gases are dissociated
in two radio-frequency dissociation sources,
with dissociation efficiency of 80-90\%, 
and are sent into
the sample chamber via two triple 
differentially pumped beam lines
\citep{Vidali2004}.

In the experiments reported here, we 
used beams of low
fluxes and short dosing times.
Using the standard Langmuir-Hinshelwood analysis,
plotting the total yield of HD vs. the exposure time
\citep{Biham2001},
we estimated the coverage to be a small fraction 
(a few percent) of a monolayer (ML).
This is still far from interstellar values but is within the 
regime in which results can be safely extrapolated to diffuse 
cloud conditions
\citep{Katz1999,Perets2005}. 
The interstellar dust analogues we used are amorphous silicate
samples, 
(Fe$_{x}$, Mg$_{1-x}$)$_{2}$SiO$_{4}$, 
prepared by one of us (J.R.B.) by laser ablation
(wavelength 266 nm) of a mixed MgO, FeO and SiO$_{2}$ target in an
oxygen atmosphere (10 mbar). 
The optical and stochiometric characterization
of samples produced with this technique is given elsewhere 
\citep{Brucato2002}. 
The results reported here are 
for a sample with $x=0.5$.

The experiment consists of adsorbing hydrogen atoms onto the surface
while monitoring the amount of hydrogen molecules that are
formed. 
To increase the signal to noise ratio, hydrogen and deuterium
atoms are used and the formation of HD is monitored. 
The measurement of HD formation is done in two steps. 
First, we record the amount of HD that forms and comes off 
the surface while the sample is being dosed with H and D atoms 
(the \emph{irradiation phase}). 
Next, after dosing is completed, in a  
TPD experiment, the surface temperature
is raised rapidly and the rate of HD desorption
is measured
(the \emph{TPD phase}). 
By far, the main contribution comes from the TPD phase.

Irradiations with beams of H and D 
(\char`\"{}H+D\char`\"{} thereafter) 
were done on an amorphous silicate surface 
in order to explore the formation processes
of HD molecules.
The H+D irradiation runs were performed
with different irradiation times (15, 30, 60, 120 and 240 s),
at a surface temperature of $T_{0} \simeq 5.6$K. 
In a separate set of experiments, beams
of HD molecules were irradiated on the same surface.
During the TPD runs, the sample temperature was monitored as a function of
time. The temperature ramps $T(t)$ 
deviate from linearity but are highly reproducible
(see inset in Fig. 2).

\begin{figure}
\includegraphics[scale=0.2,angle=270, clip,width=1\columnwidth,keepaspectratio]{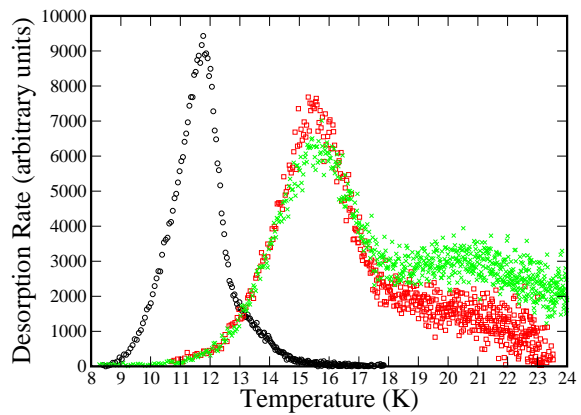}
\caption{
TPD curves of HD desorption
after irradiation of 
HD molecules ($\times$) 
and 
H+D atoms ($\square$)
on amorphous silicate. 
Also shown, for comparison, 
is HD desorption after irradiation
with H+D atoms on polycrystalline silicate ($\circ$). 
}
\label{fig:1} 
\end{figure}

\begin{figure}
\includegraphics[scale=0.2,angle=270, clip,width=1\columnwidth,keepaspectratio]{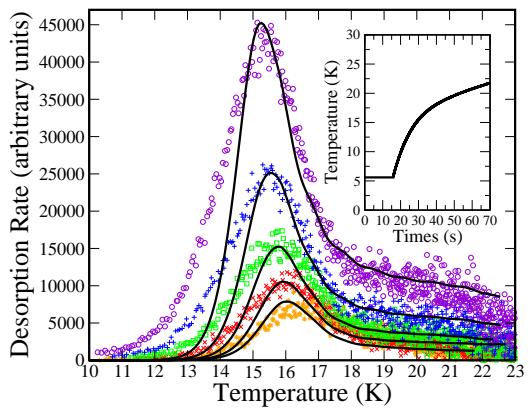}
\caption{
TPD curves of HD desorption
after irradiation with H+D atoms on amorphous silicate  with 
irradiation times of 15 ({*}), 30 ($\times$), 60 ($\Box$)
120 (+) and 240 ($\circ$) s.
The solid lines are fits
obtained using the rate equation model.
The temperature ramps (shown in the inset)
are identical for all the runs. 
}
\label{fig:2} 
\end{figure}

The desorption rates of HD molecules vs. surface temperature during the
TPD runs are shown in Fig. 1,  
for H+D irradiation on polycrystalline silicate ({\it circles})
and amorphous silicate ({\it squares})
surfaces, with irradiation times of 120 s.
The TPD curve following 
irradiation of HD molecules on an amorphous silicate 
surface
is also shown 
({\it crosses}). 
The results of current experiments of H+D irradiation on amorphous silicates,
clearly differ from those of earlier experiments
on polycrystalline silicates.
The desorption curves from amorphous silicates contain two wide peaks,
located at a significantly higher temperatures than the single narrow
peak obtained for the polycrystalline silicate.
The higher peak temperatures 
indicate that the relevant energy
barriers are larger, while their large width
reflects a broader distribution 
of the energy barriers of the HD desorption
sites. 
The TPD curve of HD desorption from amorphous silicates, after irradiation
with HD molecules (crosses in Fig. 1),
is qualitatively similar to the curve obtained for H+D irradiation.
In particular, the peak temperatures are the same.
The relative weights of the high temperature peaks vs. the
low temperature peaks are somewhat different.
Also, in similar experiments with higher values of $T_0$,
a third peak was observed at higher temperatures.
We attribute this behavior to diffusion of HD molecules, 
which gradually migrate from shallow into
deep adsorption sites 
\cite{Perets2005,Dulieu2005,Amiaud2006}.

In Fig. 2 we present a series of
TPD curves for HD desorption
after irradiation with H+D 
atoms on an amorphous silicate surface with different 
irradiation times.
Each curve exhibits a large peak at a lower temperature and
a broader peak (or a shoulder) at a higher temperature.
The position of the high temperature peak is found to be 
independent of the irradiation time,
indicating that this peak exhibits first order kinetics. 
The low temperature peak shifts to the right 
as the irradiation time is reduced.
Since in the low-coverage regime studied here,
the activation energies and the
pre-exponential factor are not expected to 
depend on the coverage,
these results indicate that the low temperature
peak exhibits second order kinetics. 
This is unlike the case of irradiation with HD molecules,
where the lower-temperature peak does not shift, 
showing first order kinetics.

\section{Analysis of the Experimental Results}

\label{sec:analysis} 

In the model
used here, there is no distinction between the H and D atoms, namely
the same diffusion and desorption barriers are used for both isotopes.
Hydrogen atoms that stick to the surface hop as random walkers
and may either encounter each other and form molecules,
or desorb from the surface. 
As the sample temperature is raised, both
the diffusion and desorption rates quickly increase. 
If a large fraction of the energy released when two H
atoms recombine 
is transformed into kinetic energy of the formed molecule, it would
immediately desorb from the grain surface in a high 
ro-vibrational state, and with large translational energy.
However, both our experiments on ice  
\citep{Perets2005}
and the current experiments
indicate that such prompt desorption does not occur on amorphous surfaces
[but see Perry and Price (2003), Tine et al. (2003) and Creighan et al. (2006)
for other, more ordered, surfaces]. 
Instead, 
the newly formed molecules dissipate their energy,
probably through multiple collisions with the rough surface or internal
pores. 
These molecules thermalize with the surface and become trapped in adsorption
sites before they thermally desorb.
Consequently, the desorbed
molecules are not highly excited, and desorb only with a thermal energy
comparable with the grain surface temperature. 
Therefore, in our model
it is assumed that the newly formed molecules do not promptly
desorb, but are trapped in adsorption sites with a range of potential
barriers. 

The experimental results were fitted using
the rate equation model described in 
Perets et al. (2005).
The parameters for the diffusion
and desorption of hydrogen atoms and molecules 
on the amorphous silicate surface were obtained. 
These include the energy barrier 
$E_{\rm H}^{\rm diff}$
for the diffusion of H atoms
and the barrier
$E_{\rm H}^{\rm des}$  
for their desorption.
The value obtained for the desorption barrier
should be considered only as a lower bound, 
because the TPD results are insensitive 
to variations in 
$E_{\rm H}^{\rm des}$,  
as long as it is higher than the reported value.
The desorption barriers of HD molecules adsorbed
in shallow (lower temperature peak)
and deep (higher temperature peak) 
sites, are given by 
$E_{\rm H_{2}}^{\rm des}(j),$ 
where $j=1$, 2, respectively.

The rate equation model is integrated using a Runge-Kutta stepper. 
For any given choice of the parameters, one obtains a set of TPD
curves for the different irradiation times used in the experiments.
The actual temperature curve of the sample, recorded during
the experiment, is used in the analysis
(see inset in Fig. 2).
In the first step, the barriers 
$E_{{\rm H_{2}}}^{{\rm des}}(j)$, $j=1$, 2, 
for the desorption of molecules
are obtained using the results of the experiments in which
HD molecules are irradiated on the surface.
To obtain better fits, we incorporate suitable 
Gaussian distributions of energy barriers 
around these two values.
In the second step, the barriers for diffusion and desorption
of H atoms are obtained,  
using the model to fit the results of 
H+D irradiation experiments
(Table 1).

\begin{table}
\caption{
Energy barriers for diffusion and desorption.
}
\vspace{0.1in}
\begin{centering}
\begin{tabular}{lccc}
\hline 
Material &
$E_{\rm H}^{\rm diff}$(meV) &
$E_{\rm H}^{\rm des}$(meV)  &
$E_{\rm HD}^{\rm des}$(meV)
\\
\hline 
Polycrystalline Silicate&
25&
32 &
27 \\
Amorphous Silicate&
35&
44&
35, 53 \\
\hline
\end{tabular}
\par
\end{centering}
\label{t:E_barriers} 
\end{table}

The second order behavior of the low temperature 
peak in the H+D irradiation
experiments can be explained as follows.
Most HD molecules are formed only when the surface temperature
is sufficiently high to enable significant mobility of H and D
atoms.
At this temperature, the shallow adsorption sites
cannot retain the newly formed molecules adsorbed in 
these sites, which quickly desorb from the surface
by thermal activation. 
However, those newly formed
molecules which are trapped in deeper sites 
do not desorb yet, and remain on the surface
until its temperature increases further.
Thus, the high temperature peak exhibits 
first order kinetics and is located
at the same temperatures in both HD and H+D irradiation experiments.
Note that we cannot exclude the possibility that some fraction
of the molecules desorbed in the high temperature peak
are formed during irradiation and remain trapped
in deep adsorption sites until the temperature becomes sufficiently
high for them to desorb.

At longer H+D irradiation times, more atoms are adsorbed on the 
surface, and they find each other more easily. 
The low temperature peak 
(corresponding to HD desorption from shallower sites) shifts
to lower temperatures with longer irradation times, 
thus showing second order behavior.
This behavior must saturate with even longer irradiation times, 
when the temperature needed for HD formation becomes
lower than the temperature needed for thermal desorption from the
shallower HD adsorption sites. 
The HD molecules then form at
low temperatures, but are trapped in the shallow sites. 
They then await until the temperature is further increased,
much like the adsorbed HD molecules in the HD irradiation experiments.
Therfore, at longer irradiation times the 
low temperature peak should saturate
into first-order like behavior and the TPD curves of the H+D and HD
irradiation experiments should become more similar. This behavior
is confirmed by the TPD curves 
for H+D and HD irradiation shown in
Fig. 1. 

Using the parameters obtained from the experiments we now calculate
the recombination efficiency of hydrogen on 
amorphous silicate surfaces
under interstellar conditions. 
The recombination efficiency is defined
as the fraction of hydrogen atoms adsorbed on the surface which come
out as molecules. 
In Fig. 3 we present the recombination efficiency
vs. surface temperature for 
the amorphous silicate sample under flux of
5.2 $\times$ $10^{-10}$ (ML s$^{-1}$). 
This flux is within the typical
range for diffuse interstellar clouds, 
where bare amorphous silicate grains are expected to play a crucial role
in H$_2$ formation. 
This flux
corresponds to gas density of 10 
(atoms cm$^{-3}$), 
gas temperature of 100K and 
density of
$7 \times 10^{14}$ adsorption
sites per cm$^{2}$ on the surface, 
obtained using the procedure
described in Biham et al. (2001). 
\begin{figure}
\includegraphics[scale=0.3, clip]{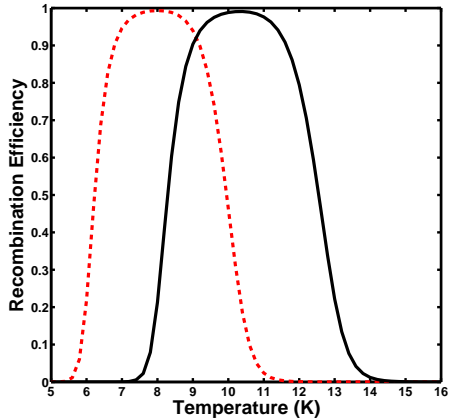}
\caption{
Calculated recombination efficiency of Hydrogen  at steady
state on amorphous silicate  (solid line) and polycrystalline silicate  (dashed line) vs.
temperature, using the parameters obtained from 
the TPD experiments.
}
\label{fig:3} 
\end{figure}

A window of high recombination efficiency is found between 8-13K,
compared to 6-10K for polycrystalline silicate under similar conditions. 
For gas density of 100 
(atoms cm$^{-2}$),
the high efficiency window for the amorphous silicates 
surface shifts to 9-14K.
At higher temperatures
atoms desorb from the surface before they have sufficient time to
encounter each other. 
At lower temperatures diffusion is suppressed and
the Langmuir-Hinshelwood mechanism is no longer efficient.
Saturation of the surface with immobile
H atoms 
might render the Eley-Rideal mechanism more efficient
in producing some 
recombination 
\citep{Katz1999,Perets2005}. 
Our results
thus indicate that recombination efficiency of hydrogen on 
amorphous silicates is high
in this temperature range, which is relevant to interstellar clouds.
Therefore, amorphous silicates 
seem to be good candidates for interstellar
grain components on which hydrogen recombines 
with high efficiency.

\section{Discussion and Summary}

\label{sec:discussion}

The analysis of the TPD curves from amorphous silicate surfaces 
shows that the relevant energy barriers on these
surfaces are significantly higher
than on polycrystalline silicates 
\cite{Katz1999,Cazaux2004}.
A similar trend was observed in amorphous and porous
ice surfaces 
\cite{Williams2007}.
These results confirm the effect of surface
morphology on the distribution of energy barriers. 
This effect can be parameterized using a model
that provides a quantitative connection between
the roughness and the energy barriers
\citep{Cuppen2005}.
Our results are consistent with this model
and can be used to quantitatively constrain 
its parameters. 
More specifically, 
we find a 1.4-1.5 times increase in
the energy barriers of the amorphous 
silicate vs. the polycrystalline silicate surface.
This gives rise to shifting and 
broadening of the temperature window in which
H$_2$ formation is efficient, by a similar factor.
Scanning electron microscope images show that the morphology of the
amorphous silicate samples is very rough
(H. Cuppen, private communication).
These samples are made of a broad (log-normal like) distribution 
of spheres and agglomerates, from
several nanometers up to a micron size. 
It is safe to conclude that 
our present results, together with those on amorphous carbon
\citep{Pirronello1999,Katz1999},
show that amorphous silicate and carbon grains
are efficient catalysts 
for the formation of molecular hydrogen in diffuse clouds.
However, the results indicate that
surface roughness
is unlikely to extend the window of efficient recombination 
to temperatures of the order of
30-50K observed in photo-dissociation regions (PDRs). 
We thus conclude that surface roughness, by itself, 
does not explain the high abundance of H$_{2}$ in PDRs.

In the model used here, it is assumed 
that H$_{2}$ molecules do not desorb immediately upon formation. 
Instead, they stay trapped in the adsorption sites or hop 
between them until thermal desorption takes place. 
Consequently, one needs to consider
mechanisms for the dissipation of the excess energy acquired from
the recombination process in order to prevent prompt desorption. 
For amorphous, porous ice surfaces it was shown, 
using time of flight (TOF) measurements, 
that the kinetic energy of the desorbed molecules is small 
($\simeq$3 meV), namely
the excess energy is absorbed by the surface
\cite{Roser2003,Hornekaer2003,Hornekaer2005}. 
Although we do not have direct measurements of the TOF of the HD molecules 
desorped from the amorphous silicate surface, 
the similarity between the TPD curves obtained after 
HD and H+D irradiations
indicates that newly formed HD molecules 
reside and then desorb from the same adsorption sites as
HD molecules irradiated on the surface.
In light of the
results on both ice and amorphous silicates, 
it is likely than H$_{2}$ molecules formed on realistic
interstellar dust would have low kinetic energy and would probably
not occupy excited vibrational or rotational states.

In summary, we have analyzed a set of TPD experiments on molecular
hydrogen formation and desorption from 
amorphous silicate surfaces under
conditions relevant to interstellar clouds. 
Fitting the TPD curves by
rate equation models, the essential parameters of 
H$_{2}$ formation on amorphous silicate surfaces were obtained. 
These parameters include the energy barrier for 
diffusion of H atoms
as well as their barrier for desorption
(considered as a lower bound).
The distribution of barriers for desorption of H$_2$
molecules is also obtained.
Interestingly, 
a single type of adsorption site for hydrogen atoms is identified, 
vs. two types of sites for molecules. 
The fraction of the adsorption sites, which belong to each of the
two types is also evaluated. 
The rate equation model provides a unified description
of several first and second order processes. 
It enables us to extrapolate the production rate
of H$_{2}$ molecules from laboratory conditions to astrophysical
conditions. 
It thus provides a quantitative evaluation of the efficiency
of amorphous silicate surfaces as 
catalysts in the formation of H$_{2}$ molecules
in interstellar clouds. 
We find that the recombination efficiency
strongly depends on the surface temperature. 
In particular, 
the amorphous silicate sample studied
here exhibits high efficiency within a range of surface temperatures
which is relevant to diffuse interstellar clouds.
The comparison of the current results with earlier ones 
on polycrystalline silicate surfaces shows the 
importance of surface morphology 
in molecular hydrogen formation.
The results are in agreement with theoretical
predictions 
on the effects of surface roughness
\citep{Cuppen2005}.
The results also indicate
that on amorphous surfaces, 
newly formed H$_{2}$ molecules are thermalized on the surface
and do not promptly desorb. 
Consequently, H$_{2}$ molecules formed
on and desorbed from realistic 
\emph{amorphous} interstellar dust
are expected to 
have very low kinetic energy and 
would probably not occupy excited
vibrational or rotational states.

This work was supported by the Adler
Foundation for Space Research and the Israel Science Foundation (O.B),
by NASA grants NAG5-11438 and NAG5-9093 (G.V), and by the
Italian Ministry for University and Scientific Research grant
21043088 (V.P).

\end{document}